\documentclass[twocolumn,trackchanges]{aastex631}

\usepackage{color}

\shorttitle{The LMC globular cluster NGC~2005}
\shortauthors{A.E. Piatti and Y. Hirai}

\begin{document}

\title{The Origin of the Large Magellanic Cloud Globular Cluster NGC~2005}

\author[0000-0002-8679-0589]{Andr\'es E. Piatti}
\affiliation{Instituto Interdisciplinario de Ciencias B\'asicas (ICB), CONICET-UNCUYO, Padre J. Contreras 1300, M5502JMA, Mendoza, Argentina}
\affiliation{Consejo Nacional de Investigaciones Cient\'{\i}ficas y T\'ecnicas (CONICET), Godoy Cruz 2290, C1425FQB,  Buenos Aires, Argentina}

\author[0000-0002-5661-033X]{Yutaka Hirai}
\altaffiliation{JSPS Research Fellow}
\affiliation{Department of Physics and Astronomy, University of Notre Dame,
225 Nieuwland Science Hall, Notre Dame, IN 46556, USA}
\affiliation{Astronomical Institute, Tohoku University,
6-3 Aoba, Aramaki, Aoba-ku, Sendai, Miyagi 980-8578, Japan}
\affiliation{Joint Institute for Nuclear Astrophysics, Center for the Evolution of the Elements (JINA-CEE), USA}

\correspondingauthor{Andr\'es E. Piatti}
\email{e-mail: andres.piatti@unc.edu.ar}

\begin{abstract}

The ancient Large Magellanic Cloud (LMC) globular cluster NGC~2005 has recently been reported
to have an \textit{ex-situ} origin, thus, setting precedents that the LMC could have partially
formed from smaller merged dwarf galaxies. We here provide additional arguments from which
we conclude that is also fairly plausible an \textit{in-situ} origin of NGC~2005, based on the
abundance spread of a variety of chemical elements measured in dwarf galaxies,
their minimum mass in order to form globular
clusters, the globular cluster formation imprints kept in their kinematics, and the recent
modeling showing that explosions of supernovae are responsible for the observed chemical abundance
spread in dwarf galaxies. The present analysis
points to the need for further development of numerical simulations and observational indices that can 
help us to differentiate between two mechanisms of galaxy formation for the LMC, namely, a primordial
dwarf or an initial merging event of smaller dwarfs.

\end{abstract}


\section{Introduction} \label{sec:intro}
\citet{mucciarellietal2021} recently reported that the Large Magellanic Cloud (LMC) old
globular cluster NGC~2005 is the unique surviving relic of a low star formation efficiency
dwarf galaxy that merged into the LMC in the past. The \textit{ex-situ} origin of NGC~2005 was 
claimed from its deficient abundance of some chemical species -- forming from different
nucleosynthesis channels -- with respect to those of five LMC old globular clusters of 
similar metallicities ($-$1.75 $<$ [Fe/H]\footnote{ {[A/B] = $\log_{10}({N_{\mathrm{A}}}/{N_{\mathrm{B}}})-\log_{10}({N_{\mathrm{A}}}/{N_{\mathrm{B}}})_{\odot}$, where $N_{\mathrm{A}}$ and 
$N_{\mathrm{B}}$ are the number densities of elements A and B, respectively.}} $<$ $-$1.69). 
A Fornax-like progenitor of 
NGC~2005 was suggested by arguing that such a dwarf spheroidal galaxy matches the 
peculiar chemical composition of NGC~2005. NGC~2005 is a 13.77$\pm$4.90 Gyr old 
globular cluster \citep{wagnerkaiseretal2017}, with a overall metallicity [Fe/H] =
 $-$1.75$\pm$0.10 \citep{setal92,beetal2002,mucciarellietal2010}, and a total mass of
  log($M/M_{\odot}$) =   5.49$\pm$0.16 \citep{mg2003}.

While the proposed scenario for the formation of NGC~2005 results plausible in the context 
of the hierarchical assembly of galaxies according to the standard cosmological model
\citep[e.g.,][]{mooreetal1999}, there are a couple of inferences made by \citet{mucciarellietal2021} 
in order to conclude on the \textit{ex-situ} origin of NGC~2005 that may allow another interpretation.

Precisely, this work aims to introduce them so that they can trigger further analysis.
The arguments in this work imply that the \citet{mucciarellietal2021}' results would not be conclusive but greatly enrich 
the debate on the NGC~2005 origin. The possible \textit{in-situ} or \textit{ex-situ} formation of NGC~2005 points to the need for a better
understanding of galaxy formation. Particularly, whether the LMC partly formed through
the accretion of smaller galactic systems or from a purely gaseous outside-in formation scenario
\citep{carreraetal2011,pg13} is still under debate. Furthermore, the analysis of the 
origin of NGC~2005 can shed light on some distinctive features that an LMC-like galaxy 
formed as a primordial dwarf should have with respect to an LMC-like galaxy partially built 
from the accretion and merging of smaller sub-units. In this context, it is worth studying
whether there is a minimum mass budget to differentiate the above two modes of 
galaxy formation. As far as we are aware, there are no simulations testing whether it is
possible to distinguish both modes of galaxy formation.


In this work, we gathered relevant works available in the literature about
mechanisms of nucleosynthesis that take place during the early life of galaxies in 
order to show that there exists an alternative interpretation for the origin of NGC~2005 
to that suggested by \citet{mucciarellietal2021}. The present results do not discredit 
the possible \textit{ex-situ} origin of this ancient LMC globular cluster but pose the issue in a 
broader context. These results point to the need for detailed simulations exploring
the space of similarities and differences of galaxy formation processes for different
galaxy masses. In order to provide a conclusive answer about the origin
of NGC~2005, further spectroscopic observation campaigns of LMC field stars are needed, as well
as globular cluster formation modeling in the context of galaxy formation with higher resolution and 
precision of the orbital integration. Nevertheless, the present somehow qualitative
arguments shed light on a more comprehensive analysis of the origin of NGC~2005.

\section{Analysis}

The first piece of analysis that led us to support a possible \textit{in-situ} origin of NGC~2005 is
a  rigorous statistical treatment  of the abundances used by \citet{mucciarellietal2021} 
to conclude on the \textit{ex-situ} origin.
\citet{mucciarellietal2021} showed that 13 chemical abundances measured in NGC~2005
are systematically lower  than the values for five LMC globular clusters 
(NGC~1786, 1835, 1916, 2210, 2257)  with metallicities ([Fe/H])  similar to that of NGC~2005. 
However, if the uncertainties are taken into account, those differences change as a function
of the chemical element.

\begin{figure}
\includegraphics[width=\columnwidth, bb = 0 0 450 350]{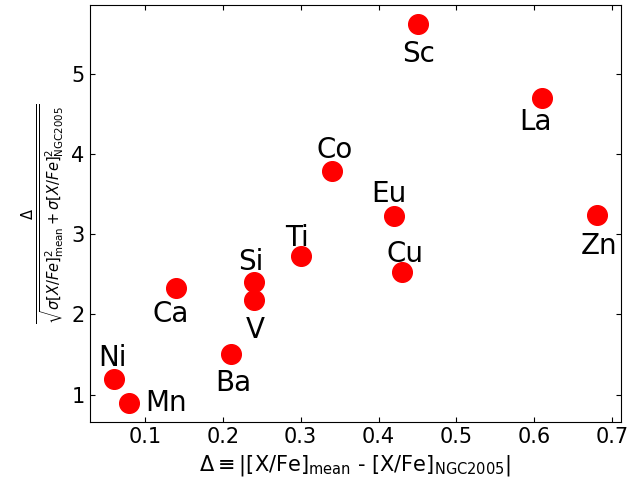}
\caption{Diagnostic diagram built to illustrate the quality of the measurements
of different chemical abundances ratios in \citet{mucciarellietal2021}.}
\label{fig1}
\end{figure}

Figure~\ref{fig1} shows a more comprehensive picture in this respect. 
In order to build it, 
we first computed the difference ($\Delta$, absolute value) in abundance ratios between the mean 
abundance ratios for the aforementioned five LMC globular clusters ([X/Fe]$_{\rm mean}$) and 
that of NGC~2005 ([X/Fe]$_{\rm NGC2005}$),
using values kindly provided by A. Mucciarelli. We note that \citet{mucciarellietal2021}
only included the values for [Si/Fe], [Ca/Fe], [Cu/Fe], and [Zn/Fe] in their table~1,  because they 
focused on elements with predictions of stellar yields that are representative of different 
nucleosynthesis channels. Particularly, they rely their analysis on the [Zn/Fe] ratio, for which they
found the largest mean difference (0.68 dex).

For completeness purposes, we included in
Table~\ref{tab1} all the [X/Fe] ratios used in that work. Then, we added  in quadrature their
respective uncertainties $\sigma$[X/Fe]$_{\rm mean}$ and $\sigma$[X/Fe]$_{\rm NGC2005}$,
and calculated $\eta$=$\Delta$/$\sqrt{\sigma [\rm{X/Fe}]_{\rm mean}^2+\sigma [\rm{X/Fe}]_{\rm NGC2005}^2}$, which
 we plotted in Figure~\ref{fig1} as a function of $\Delta$, and included them in Table~\ref{tab1}. 
As can be seen,  [Sc/Fe],  [Co/Fe], [La/Fe], [Zn/Fe], and [Eu/Fe]  ratios show differences $\Delta$ larger than 
3 times the sum of their respective uncertainties, which means these chemical element abundances in 
NGC~2005 and the other five globular clusters are different. We note that $\eta$ values $<$ 3 do not 
warrant a real difference so that
the conclusion on distinctive chemical patterns between NGC~2005 and the other five LMC
globular clusters for  [Si/Fe], [Ca/Fe], and several other chemical abundances should be taken with 
caution. Therefore, if only some chemical elements show real abundance differences between NGC~2005
and 5 LMC globular clusters, both \textit{ex-situ} and \textit{in-situ} formation scenarios are feasible.

\begin{deluxetable}{lccccc}
\tablecaption{[X/Fe] values from \citet[private communication]{mucciarellietal2021}.}
\label{tab1}
\tablewidth{0pt}
\tablehead{\colhead{[X/Fe]} & \colhead{NGC~2005} & \colhead{$<$LMC$>$} & \colhead{$\Delta$} & \colhead{$\sigma$} & \colhead{$\eta$}}
 \startdata
 Si	& 0.08$\pm$0.09 & 0.32$\pm$0.05 & 0.24 & 0.10 & 2.40 \\
 Ca &  0.01$\pm$0.05 & 0.15$\pm$0.03 & 0.14 & 0.06 & 2.33\\
 Sc	& $-$0.39$\pm$0.07 &0.06$\pm$0.04 & 0.45 & 0.08 & 5.62 \\
Ti	& $-$0.06$\pm$0.10 & 0.24$\pm$0.04 & 0.30 & 0.11 & 2.73 \\
V	& $-$0.34$\pm$0.10 & $-$0.10$\pm$0.04 & 0.24 & 0.11 & 2.18\\
Mn	& $-$0.61$\pm$0.09 & $-$0.53$\pm$0.02 &0.08 & 0.09 & 0.89 \\
Co	& $-$0.29$\pm$0.08 & $ $0.05$\pm$0.03 & 0.34 & 0.09 & 3.78 \\
Ni	& $-$0.07$\pm$0.05 & $-$0.01$\pm$0.02 & 0.06 & 0.05 & 1.20 \\
Cu	& $-$1.10$\pm$0.14 & $-$0.67$\pm$0.09 & 0.43 & 0.17& 2.53 \\
Zn	& $-$0.80$\pm$0.20 & $-$0.12$\pm$0.07 & 0.68 & 0.21 & 3.24 \\
Ba	& 0.09$\pm$0.09 & 0.30$\pm$0.11 & 0.21 & 0.14 & 1.50 \\
La	& $-$0.22$\pm$0.07 & 0.39$\pm$0.11 & 0.61 & 0.13 & 4.69 \\
Eu	& 0.28$\pm$0.06 & 0.70$\pm$0.11 & 0.42 & 0.13 & 3.23 \\
\enddata
\end{deluxetable}

Figure~\ref{fig1} shows that  the 13 different chemical abundances in 
NGC~2005 and the other five LMC  clusters would not seem to be equally distinguishable, 
and those differences are only exhibited for a couple of the 13  chemical species analyzed
by \citet{mucciarellietal2021}. The unavoidable question arises: what does the difference
in these few chemical species mean? \citet{mucciarellietal2021}
argued that NGC~2005 formed in a Fornax-like dwarf galaxy that was accreted onto the LMC.
According to them, the  Fornax-like dwarf galaxy would have left negligible consequences
in the LMC in the form of relics (galaxy mass ratio $<$ 0.01), except only NGC~2005.
The different chemical abundance features would be a signature that NGC~2005 formed
\textit{ex-situ} the LMC.

\begin{figure}
\plotone{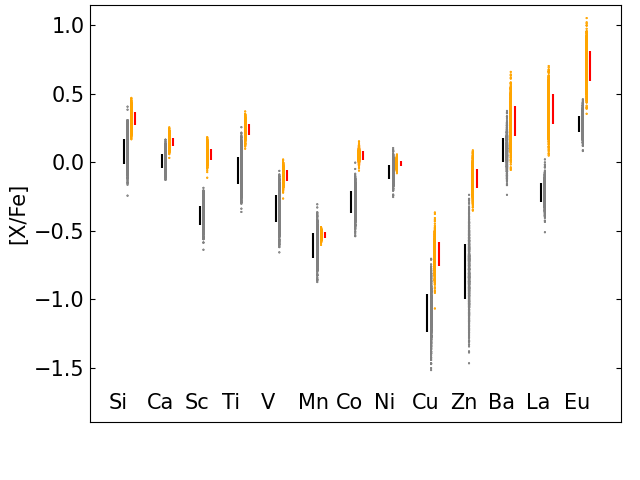}
\caption{[X/Fe] ratios derived by \citet{mucciarellietal2021}
(see Table~\ref{tab1}) for NGC~2005 (grey points) and the
other five LMC globular clusters (orange dots), respectively. Each [X/Fe] ratio is
represented by 1000 points following a gaussian distribution. For the seek of the
reader we included the corresponding 1$\sigma$ error bars represented by black
and red segments, respectively.}
\label{fig2}
\end{figure}

However, the global chemical pattern of NGC~2005  is not statistically peculiar 
compared to that 
of the LMC. In order to quantify this, we figured out that we measured 1000 times the
abundance of each of the 13 elements  of Table~\ref{tab1}  in NGC~2005 and in the other 
five LMC globular clusters; then we gathered the 1000 measurements of each element,
and looked at the obtained distributions. We assume, as expected, that this experimental 
exercise will provide normal distribution functions, so that we represented them by
generating 1000 points following a gaussian distribution for each element in NGC~2005
and in the other five LMC globular clusters. In order to do this, we used the {\it random.normal}
library within Numpy\footnote{\url{https://www.numpy.org/}} using the mean values and
errors quoted in Table~\ref{tab1} as {\it loc} and {\it scale} parameter values, respectively
(see Figure~\ref{fig2}).  Note that most of the points are concentrated within 1$\sigma$.
As can be seen, there
are some elements in NGC~2005 and in the five LMC globular clusters
whose point distributions totally overlap (Mn, Ni); other elements with 
a partial overlap (e.g., Ca, V, Ba), and a few ones which look different (Sc, La).
The total overlap of the point distributions for some chemical elements means that 
any possible measure of that element in NGC~2005 can also be obtained for the other
five LMC globular clusters, or role reversal. A similar reasoning can be used for those
elements with a partial or null overlap, respectively. We note that the comparison of these
distributions for each element in NGC~2005 and in the five LMC globular clusters
is more meaningful than the use of the respective $\eta$ 
values (see Table~\ref{tab1}),  although the latter has the advantage of providing
a quantitative measure.

 If we considered altogether the 13 element distributions of NGC~2005 
and compared it with that of the five LMC globular clusters, we would obtain
a measure of the level of similitude between them.
We then
considered the 13 chemical elements together using the 26000 points of Figure~\ref{fig2};
 most of the points distributed within 1$\sigma$ as provided by the normal distribution
law. We built two tables, one for NGC~2005 and another for the five LMC globular clusters
containing the respective 13000 generated previously. Then we statistically estimated the similarity
between these two tables - in a scale from 0 to 1, where 0 means totally different and 1 
means totally equals - between
the 13 chemical abundances in NGC~2005  (grey points in Fig.~\ref{fig2}) and the LMC 
 (orange points in Fig.~\ref{fig2}) 
 using different statistical 
methods, namely: Jaccard similarity (0.47); cosine distance (0.59); S{\o}rensen-Dice statistic (0.64);
Levenshtein, Hamming, Jaro, and Jaro-Winkler distances (0.52); Pearson correlation (0.83); Spearman
correlation (0.81). We used  Python language v3.8.10\footnote{\url{http://www.python.org/}}, 
and the following packages: Numpy; 
Scipy\footnote{\url{https://scipy.org}}; and PyPi\footnote{\url{https://pypi.org/}}, also found in GitHub 
repository\footnote{\url{https://github.com/}}. The Appendix shows the Python scripts used
in order to applied the aforementioned statistics. The resulting similarities between both
samples are given within parenthesis above, following the name of the respective method.
As can be seen, the general consensus of these
statistics is that the chemical abundance pattern of NGC~2005 can partially overlap 
that of the other 5 LMC globular clusters.

In order to explore such a possibility more deeply, we first thoroughly searched the literature for [X/Fe] ratios measured in the LMC. \citet{hasselquistetal2021} used APOGEE abundances 
\citep{majewskietal2017} to uncover the chemical abundance patterns in massive Milky Way satellites, including the LMC. They carefully selected galaxy stars based
on APOGEE data quality cuts and proper motions. 
For Si, Ca, and Ni (the only three 
elements overlapping those in \citet{mucciarellietal2021}), they obtained the median abundances
and $\pm$1 $\sigma$ uncertainties listed in Table~\ref{tab2}. When comparing these values
with those of five LMC globular clusters in Table~\ref{tab1}, we found that abundances
of Ca and Ni are  less than 1$\eta$ and that of Si is different at 2.5$\eta$ level.
We assume that this result supports that both metallicity scales are similar within the quoted
uncertainties, inhomogeneities and/or systematic effects, if present, being smaller.
We carried out the same statistical
comparison for the values of NGC ~2005,  and found that abundances of Si and Ni are less 
than 1$\eta$, but Ca is different at 1.6$\eta$. They suggest that these chemical elements in
NGC~2005 and the LMC have similar abundances.

\begin{deluxetable*}{lcccccccc}
\tablewidth{0pt}
\tablecaption{Mean [X/Fe] values in dex at the NGC~2005 metallicity ([Fe/H]=$-$1.75$\pm$0.10 dex)
for the observed [X/Fe] vs. [Fe/H] distributions  (References in parenthesis).}
\label{tab2}
\tablehead{\colhead{X}	       &  \colhead{LMC}		& \colhead{Fornax} 			&\colhead{Sagittarius} 		&	\colhead{Sextants} 		& \colhead{Sculptor} 		& \colhead{Ursa Minor}}
\startdata
Si	&0.17$\pm$0.03 (1) 	&				&0.17$\pm$0.04 (1)	&				& 	&	\\
Ca	&0.12$\pm$0.05 (1) 	&				&0.02$\pm$0.07 (1)	&				&	&	\\
Sc	&	 	 		&$-$0.07$\pm$0.15 (2)	&$-$0.10$\pm$0.02 (2)	&$-$0.28$\pm$0.17 (2)	&0.00$\pm$0.10 (2)	&$-$0.30$\pm$0.10 (2)\\
Ti	&	 			&0.25$\pm$0.10 (2)	&0.25$\pm$0.10 (2)	&0.25$\pm$0.15 (2)	&0.20$\pm$0.25 (2)	&0.25$\pm$0.15 (2)	\\
V	&	 			&				&				&				&				&	\\
Mn	&	 			&$-$0.60$\pm$0.10 (2)	&$-$0.75$\pm$0.15 (2)	&$-$0.55$\pm$0.10 (2) &$-$0.35$\pm$0.10 (2)&$-$0.70$\pm$0.05 (2)	\\
Co	&	 			&				&				&				&				&	\\
Ni	&$-$0.03$\pm$0.03 (1) &0.00$\pm$0.05 (2)	&$-$0.03$\pm$0.04 (1) & 0.00$\pm$0.10 (2)&$-$0.10$\pm$0.13 (2)  &$-$0.10$\pm$0.07 (2)	\\
	&				&				&$-$0.05$\pm$0.06 (2)&				&				&				\\
Cu	&	 			&				&				&				&				&	\\
Zn	&	 			&				&$-$0.03$\pm$0.12 (2)	&				&$-$0.25$\pm$0.20 (2) &$-$0.20$\pm$0.19 (2)	\\
Ba	&	 			&$-$0.10$\pm$0.07 (2)	&0.01$\pm$0.04 (2)	&0.00$\pm$0.20 (2)	&$-$0.10$\pm$0.20 (2) &$-$0.05$\pm$0.15 (2)	\\
La	&	 			&				&				&				&	&	\\
Eu	&				&				&0.50$\pm$0.05 (2)	&0.55$\pm$0.05 (2)	&0.45$\pm$0.04 (2)&0.50$\pm$0.04 (2)	\\
\enddata
\tablecomments{References: (1) \citet{hasselquistetal2021}; (2) \citet{reichertetal2020}.}
\end{deluxetable*}

We also derived the mean values of different chemical elements of dwarf
galaxies using the homogenized
analysis carried out by \citet{reichertetal2020}, which is, as far as we are aware, the largest 
compilation of these quantities for this type of object. The calculated values are listed in Table~\ref{tab2}. 
As can be seen, the abundance of Ni in Sagittarius is in excellent agreement
with that of \citet{hasselquistetal2021} so that we assumed that both results are in the same
scale within the quoted uncertainties.

For the chemical elements showing $\eta$$>$ 3 
in Figure~\ref{fig1} (Sc, Zn, and Eu), 
we repeated the statistical analysis described above  by comparing the values in Table~\ref{tab1}
with those in Table~\ref{tab2}. We found that: 1) the 5 LMC globular
clusters have  $\eta <$ 3 for Sc, Zn, and Eu with respect to all the
dwarfs included in Table~\ref{tab2},  with the exception of Sc for Sagittarius and Ursa Minor; 2) 
NGC~2005 has only Sc abundance different from Sagittarius and Sculptor and Zn abundance 
different from Sagittarius. The above results show that chemical element abundances that appear 
to be different in NGC~2005 with respect to LMC are also found to be different, unevenly, 
between the LMC and other dwarfs, as well as between NGC~2005 and other dwarfs.


 Note that we performed this statistical analysis based on chemical abundances derived by the different sets of analysis, i.e., we adopt chemical abundances of \citet{mucciarellietal2021} and \citet{hasselquistetal2021}. On the other hand, \citet{mucciarellietal2021}' chemical abundances were derived with the same analysis. This difference may introduce additional inhomogeneity and systematic effects.

We would expect that if NGC~2005 formed in a dwarf (Fornax-like) galaxy  as proposed by \citet{mucciarellietal2021}, their chemical abundance patterns 
should be similar.  \citet{mucciarellietal2021} compiled from the literature 
abundances for Si, Ca, Cu, and Zn in Fornax (see their supplementary figure 5). Unfortunately, none of these chemical elements are in
the compilation by \citet{reichertetal2020} to compare one to the other. Nevertheless,
we found that \citet{letarteetal2006} measured Ba, Ni, and Ti abundances for Fornax's globular
clusters. When extrapolating their [X/Fe] vs. [Fe/H] relationships up to [Fe/H] = $-$1.75, we
found that the values for Ba and Ni are similar to those in Table~\ref{tab2}, while that
for Ti is somewhat different. Therefore, according to the compilation by \citet{reichertetal2020},
the chemical element abundances in NGC~2005 and Fornax would not seem to be
clearly different. We note that it would be worth performing further measurements for more 
chemical elements in the LMC and other dwarfs to make a more comprehensive
comparison with the values obtained for NGC~2005.

\section{Discussion}
In this study, we explore whether a scenario of \textit{in-situ} formation of NGC 2005 is still allowed. We have shown that most elements show $\eta$ between 2 and 3, meaning that the probability that the difference in 
chemical abundance is real is more than 95$\%$, it can not be considered insignificant. 
Although this argument would suffer from systematics in different observations, the following discussion of chemo-dynamical properties of NGC 2005 could support the \textit{in-situ} formation scenario.
\subsection{Chemical Abundances}
The analysis of
chemical abundances and their production channels support a possible \textit{in-situ} origin of NGC~2005 even if chemical abundances in NGC 2005 and other LMC's globular clusters are different.
Dispersion of the [Zn/Fe] ratios in dwarf galaxies can be caused by the inhomogeneity of the interstellar medium. \citet{hiraietal2018} performed a series of chemo-dynamical simulations of dwarf galaxies with the Zn enrichment. Their models assume that Zn is synthesized by electron-capture supernovae and hypernovae, while Fe is from core-collapse supernovae (CCSNe) and type Ia supernovae (SNe Ia). They found that scatters of [Zn/Fe] for [Fe/H] $> -$2.5 reflect the inhomogeneity of [Zn/Fe] ratios caused by SNe Ia. As shown in their figures 9 and 11, several stars with [Fe/H] $> -$2 have [Zn/Fe] $< -$1. These stars are formed from gas clouds heavily enriched by SNe Ia. These results mean that inhomogeneity caused by SNe Ia could produce low [X/Fe] ratios at relatively high metallicity in dwarf galaxies.

Characteristics of the [X/Fe] in NGC~2005 suggest that it was formed from the gas cloud heavily affected by the ejecta of SNe Ia. Since they synthesize a large amount of Fe, star clusters formed in gas containing ejecta of SNe Ia tend to show low [X/Fe] ratios if these types of supernovae do not largely synthesize the element X. A notable example is [Eu/Fe], which shows $\Delta$ = 0.42. Eu is almost entirely synthesized by the $r$-process, which does not occur in SNe Ia \citep[e.g.,][]{hiraietal2015, hiraietal2017, wanajoetal2021}.

On the other hand, the difference of the [X/Fe] for elements synthesized by SNe Ia tends to be smaller. The double detonation (CSDD-L) model of sub-Chandrasekhar (sub-$M_{\rm{Ch}}$) white dwarfs in \citet{lachetal2020} synthesizes a large amount of Ca, Mn, and Ni but not much for Co, Cu, and Zn. As shown in Figure \ref{fig1}, $\Delta$ of [Ca/Fe], [Mn/Fe], and [Ni/Fe] are relatively small compared to [Co/Fe], [Cu/Fe], and [Zn/Fe]. \citet{delosreyesetal2022} found the possible contribution of sub-$M_{\rm{Ch}}$ SNe Ia to the Sculptor dwarf galaxy from [Mn/Fe] ratios. These results mean that NGC 2005 could be formed from the gas cloud heavily affected by the ejecta of SNe Ia.


The lack of well-mixed gas during the formation of the 
LMC is documented in the case of Fe by the extensive range of [Fe/H] values of the 15 LMC globular clusters 
and field stars ($-$2.0 $\leq$ [Fe/H] $\leq$ $-$1.3), all of them formed in a relatively short timescale 
\citep[$\Delta$(age) $\sim$2 Gyr;][]{pg13,pm2018,piattietal2018c}. Since SNe Ia can be occurred in $\sim$1 Gyr \citep[e.g.,][]{strolger2020}, this timescale is enough to cause SNe Ia in the progenitors of the LMC.




Among the 15 LMC globular clusters, four are metal-poorer than NGC~2005; five are of comparable
metallicity, and
other five clusters are metal-richer; the whole globular cluster population spanning the
[Fe/H] range from $-$2.0 up to $-$1.3 \citep{piattietal2019}.  
If we considered the gas cloud metallicities similarly distributed as the
metallicity distribution of the LMC globular clusters (27\% metal-poorer, 40\% similar, and 33\% metal-richer than
[Fe/H]=$-$1.75), then we would find that $\sim$ 27\%,
40\% and 33\% of the whole gathered gas cloud was metal-poorer, with similar
metallicity, and metal-richer than NGC~2005, respectively. The portion of the gas cloud out of which
the five globular clusters with metallicities ([Fe/H]) similar to that of NGC~2005 and NGC~2005 itself were formed
(40\% of the whole gas cloud),  should also have the 13 chemical elements analyzed by \citet{mucciarellietal2021}
distributed similarly as these 6 globular clusters, i.e., five sharing a similar pattern and NGC~2005 with a somewhat
different one. This means that 1/6 of that gas cloud portion (40\%/6 $\approx$ 6\% of the whole cloud, 
$\sim$12$\,\times\,$10$^6$$M_\odot$) should have had the chemical 
abundance pattern found in NGC~2005. This percentage explains that only NGC 2005 has a 
different chemical abundance from other globular clusters.

This estimate is consistent with the gas mass affected by SNe Ia around dwarf galaxies formed in a cosmological zoom-in simulation. Here we analyze the high-resolution cosmological zoom-in simulation of a Milky Way-like galaxy in \citet{hiraietal2022}. This simulation assumes the initial mass function of \citet{chabrier2003} from 0.1 $M_{\odot}$ to 100 $M_{\odot}$ with the nucleosynthesis yields of \citet{nomotoetal2013} for CCSNe and the N100 model of \citet{seitenzahl2013} for SNe Ia. They also adopt a turbulence-induced metal mixing model to compute chemical inhomogeneity correctly \citep{hiraisaitoh2017}. We pick up the most massive satellite dwarf galaxy from this simulation with a total stellar mass of 2.1$\,\times\,10^7\,M_{\odot}$ at $z$ = 0. Although this simulation does not have LMC-mass systems, this satellite is large enough to discuss the inhomogeneity caused by the SNe Ia. As shown in \citet{reichertetal2020}, Fornax dwarf spheroidal galaxy, which has a similar mass to this galaxy, also has significant variations of chemical abundances.

We then compute a gas affected by SNe Ia at the lookback time of 11.5 Gyr within the virial radius of the progenitor of this galaxy. By this analysis, we found that 2.6$\,\times\,10^7\,M_{\odot}$ of gas shows $-2\,<\,$[Fe/H]$\,<\,-1$ and [Mg/Fe]$\,<\,0$, which indicates these gas clouds are affected by SNe Ia. The total gas mass of this galaxy in this epoch is 8.8$\,\times\,10^7\,M_{\odot}$.  Since globular clusters are collisional systems, galaxy formation simulations assuming collisionless systems cannot correctly resolve the formation and evolution of globular clusters. Even though there are such numerical difficulties, this result suggests that there is enough gas around a dwarf galaxy to form a globular cluster affected by SNe Ia together with globular clusters with different chemical abundances.

In addition to the analysis of the cosmological zoom-in simulation, we estimate the star formation rates (SFRs) and the number of SNe Ia ($N_{\rm{Ia}}$) to explain the chemical abundances obtained by \citet{mucciarellietal2021}. Here we estimate the SFRs and $N_{\rm{Ia}}$ using closed box chemical evolution  model \citep{hiraisaitoh2017}. In this model, we adopt exponentially declining SFR, i.e., SFRs are proportional to exp($-t/\tau$), where $\tau\,=\,2\times10^9$ yr. The initial gas mass of this system is $5\times10^9M_{\sun}$. These values result in a model consistent with the LMC's metallicity distribution and star formation timescale. Since this model is to roughly estimate SFRs and $N_{\rm{Ia}}$, we ignore gaseous inflow and outflow. For nucleosynthesis yields, we adopt \citet{nomotoetal2013} for CCSNe and the W7 model of \citet{iwamoto1999} for SNe Ia. Since this yield set tends to overproduce Si and Ca, the resulting [Si/Fe] and [Ca/Fe] are shifted $-$0.2 dex \citep{timmesetal1995,prantzosetal2018}. This shift is done within the uncertainties of nucleosynthesis.

According to this model, SFRs for $\lesssim$1 Gyr are 1 $M_{\sun}\,\rm{yr}^{-1}$ while they are decreased to $10^{-3}\,M_{\sun}\,\rm{yr}^{-1}$ at 13.8 Gyr. The final stellar mass of this system is $3\times10^9M_{\sun}$, consistent with the stellar mass of the LMC. \citet{mucciarellietal2021} also estimated that the LMC globular clusters were formed with 1--1.5 $M_{\sun}\,\rm{yr}^{-1}$ in the early phase.

We have counted the number of type Ia supernovae in this model. When the system’s metallicity is [Fe/H] = $-$1.75, the system is polluted by $5.5\times10^5$ of SNe Ia. At this time, [Si/Fe] and [Ca/Fe] are 0.36 and 0.11, respectively. These values are consistent with the average [X/Fe] values in LMC (Table \ref{tab1}).

We further estimate $N_{\rm{Ia}}$ to explain [Si/Fe] ratios of NGC 2005 by the scenario of the local inhomogeneity. We assume that NGC 2005, with the stellar mass of $3\times10^5M_{\sun}$ was formed from the gas cloud of $5\times10^7M_{\sun}$. This assumption is based on the average star formation efficiency (0.006) of the giant molecular cloud \citep{murray2011}.  By adopting the solar system abundance of \citet{asplundetal2009}, we estimate that there are 1200 $M_{\sun}$ of Fe and Si in the cloud with [Fe/H] = $-$1.75 and [Si/Fe] = 0.32. To decrease the [Si/Fe] to the value of NGC 2005 ([Si/Fe] = 0.08), we estimate that d$M_{\rm{Fe}}$ = 860 $M_{\sun}$ of Fe should be added to the cloud. We then apply the Fe yield ($Y_{\rm{Fe}}$ = 0.75 $M_{\sun}$ for each SNIa) of the W7 model of \citet{iwamoto1999} and ignore the production of Si in SNe Ia. By dividing d$M_{\rm{Fe}}$ by $Y_{\rm{Fe}}$, $N_{\rm{Ia}}$ required to reproduce [Si/Fe] in NGC 2005 is ~1100. This number is only 0.2\% of the total number of SNe Ia in the whole region.

This result means that if there is a region around LMC enriched slightly excess in the ejecta of SNe Ia, a globular cluster with alpha-element abundance similar to NGC 2005 could be formed. Since this is a simple estimation, we ignore the increase of [Fe/H] by additional SNe Ia. We also refrain from doing this estimate on other elements due to the significant uncertainties of nucleosynthesis.


\subsection{Kinematics}
Kinematics of globular clusters let us consider an  \textit{in-situ} origin for NGC~2005.
Indeed, the kinematics of 
Milky Way globular clusters have been used to disentangle different accretion events
\citep{massarietal2019,kruijssenetal2019} since their motions have kept along their lifetimes' imprints of their origins \citep{piatti2019}. Similarly, 
\citet{bennetetal2022} performed a 6D phase-space analysis from multiple independent analysis techniques
of 31 LMC globular clusters using {\it Gaia} EDR3 \citep{gaiaetal2020b} and {\it Hubble Space Telescope} data.
They found that the system of globular clusters rotates like in a stellar disk with one-dimensional velocity dispersions of order 30 
km\,s$^{-1}$, similar to that of the LMC old stellar disk population. From these results, they argued that most, if not
all, LMC globular clusters formed through a single formation mechanism in the LMC disk, albeit their
significant dispersion in age and metallicity, any accretion signature being absent within the
involved uncertainties. Similarly to outer halo Milky Way globular clusters, which are associated with
dwarf galaxy accretion events \citep[e.g., Sagittarius, Gaia-Enceladus, Sequoia, etc.,][]{forbes2020},
the LMC halo globular cluster should be those with more chances to have an 
\textit{ex-situ} origin, but at present, there is no chemical signature hinting at it. Note that
NGC~2005 is placed in the inner LMC disk. 

Several recent studies support the \textit{in-situ} scenario.
\citet{shaoetal2021} showed that accreted globular clusters in the Milky Way and Fornax
are less centrally concentrated than those formed \textit{in-situ}. Moreover, globular clusters that
escape dwarf satellites of the Milky Way are found orbiting the latter \citep{rostamietal2022}.
\citet{piattietal2019} derived mean proper motions of the 15 LMC globular clusters, and from
existent radial velocities, they computed their velocity vectors. They found that LMC globular
clusters are distributed in two different kinematics groups, namely: those moving in the LMC disk
and others in a spherical component. Since globular clusters in both kinematic–structural components share 
similar ages and metallicities, they concluded that their origin occurred 
through a fast collapse that formed a halo and disk concurrently. NGC~2005 resulted in being 
a disk globular cluster, while among the other 5 LMC clusters, three and two are in the disk and halo,
respectively.

In addition to kinematics, the mass required to form globular clusters would support the \textit{in-situ} formation scenario. \citet{eadieetal2021} estimated a minimum
galaxy stellar mass required to form globular clusters of $\sim$10$^7$$M_\odot$ 
We found from Table~\ref{tab3} that only Sagittarius and Fornax could form globular clusters.
Carina, Draco, and Ursa Minor do not have globular clusters (either formed \textit{in-situ}
or \textit{ex-situ}). With a total galaxy stellar masses of
$\sim$10$^{5.4}$$M_\odot$ it is probable that these galaxies are primordial dwarfs, i.e.,
they did not form from the merger of smaller galaxies.

\begin{deluxetable}{lcc}
\tablewidth{0pt}
\tablecaption{Stellar mass of dwarf galaxies using the absolute $M_V$ magnitudes compiled by \citet{drlicawagneretal2020} and interpolating them in figure~5 of \citet{georgievetal2010}.}
\label{tab3}
\tablehead{\colhead{Name} 		& \colhead{$M_V$ (mag)} & \colhead{log(stellar mass /$M_\odot$)} }
\startdata
Carina		&	$-$9.43		&	5.6$^{+0.3}_{-0.2}$ \\
Draco		&	$-$8.71		&	5.3$^{+0.3}_{-0.2}$ \\
Fornax		&	$-$13.46	&	7.4$^{+0.3}_{-0.2}$ \\
Sagittarius	&    $-$13.50		&	7.4$^{+0.3}_{-0.2}$ \\
Sextants		&	$-$8.72		&	5.3$^{+0.3}_{-0.2}$ \\
Sculptor		&	$-$10.82	&	6.2$^{+0.2}_{-0.1}$ \\
Ursa Minor	&	$-$9.03		&	5.4$^{+0.3}_{-0.2}$ \\
\enddata
\end{deluxetable}
We also computed the LMC mass for a lookback time of 11.5 Gyr, when
all its globular clusters formed \citep{piattietal2019}, using the SFRs derived
by \citet{mazzietal2021} and \citet{massanaetal2022}. We obtained an LMC mass of
$\sim$10$^8$$M_\odot$. Thus, the pre-enriched gas cloud out of which the LMC globular clusters formed
could have been a gathering of smaller pieces, each with a particular chemical
enrichment history.

 Following these discussions, we anticipate that stars with chemical abundances similar to NGC 2005 would be formed if there are gas clouds with enough mass ($\sim$10$^7\,M_{\sun}$) enriched by SNe Ia larger than the other region. On the other hand, in \citet{mucciarellietal2021}’ scenario, NGC 2005 would be formed around a Fornax-like dwarf galaxy and later accreted to the LMC. Our discussion suggests that NGC 2005 could be formed \textit{in-situ} without imposing unphysical assumptions.
\section{Conclusions}
The present analysis shows that the 13 chemical elements employed by
\citet{mucciarellietal2021} to claim an \textit{ex-situ} origin of NGC~2005
are not all of the same accuracy. Consequently, they cannot be used
indistinctly to support abundance differences between the [X/Fe]
values derived for NGC~2005 and for five LMC globular clusters with 
similar metallicities. Nevertheless, the abundance differences measured 
for some chemical elements ($>$ 3$\sigma$) are yielded by 
SNe Ia. Different 
dwarf galaxies with studied chemical enrichment histories show 
abundances spread of the considered chemical elements that
encompass the mean values of NGC~2005, which means that NGC~2005
could have been born in any of these galaxies, including the LMC.
The five LMC globular clusters with [Fe/H] values similar to that
of NGC~2005 belong to the inner disk (3) and the outer halo (2) 
of the LMC, and they have similar individual [X/Fe] ratios.
The LMC globular clusters span a wide range of metallicities, and 
that range is verified from those populating the kinematically 
different disk and halo substructures, respectively \citep{piattietal2019}.  
Therefore, the presence of NGC~2005 in the inner LMC disk should not catch
our attention to differentiate it from the remaining LMC globular
cluster population. Recent modeling has also shown a widespread abundance of chemical species at the metalicity level of NGC~2005
can be produced by supernova explosions, as has also been probed
in the LMC.

\begin{acknowledgments}
We thank the referee for the thorough reading of the manuscript and
timely suggestions to improve it.
We thank Alessio Mucciarelli for
providing data for Table~\ref{tab1}, and Giuseppina Battaglia for
valuable comments. Y.H. is supported in part by JSPS KAKENHI Grant Numbers JP21J00153, JP20K14532, JP21H04499, JP21K03614, JP22H01259, MEXT as ``Program for Promoting Researches on the Supercomputer Fugaku" (Toward a unified view of the universe: from large scale structures to planets, Grant No. JPMXP1020200109), JICFuS, and grants PHY 14-30152; Physics Frontier Center/JINA Center for the Evolution of the Elements (JINA-CEE), and OISE-1927130: The International Research Network for Nuclear Astrophysics (IReNA), awarded by the U.S. National Science Foundation. Numerical computations and analysis were carried out on Cray XC50 and computers at the Center for Computational Astrophysics, the National Astronomical Observatory of Japan, and the Yukawa Institute Computer Facility. This research has made use of NASA's Astrophysics Data System Bibliographic Services.
\end{acknowledgments}



\begin{appendix} 

\section{Python scripts used to perform different statistics}

In what follows y1 and y2 represent values of the
13 element abundances in NGC~2005 and in the five LMC
globular clusters, respectively (see text for details). \\

$\#$ Spearman correlation\\

spearman= stats.spearmanr(y1,y2, axis=0)

print (spearman) \\

$\#$ S{\o}rensen-Dice statistic\\

def dice(a, b):

    $\_$a = set(a)

    $\_$b = set(b)

    return (2*len($\_$a.intersection($\_$b))) / (len($\_$a) + len($\_$b))

dice = dice(y1, y2)

print (dice) \\



$\#$ Jaccard similarity \\

def jaccard$\_$similarity(x,y):

intersection$\_$cardinality = len(set.intersection(*[set(x), set(y)]))

union$\_$cardinality = len(set.union(*[set(x), set(y)]))

return intersection$\_$cardinality/float(union$\_$cardinality)

jaccard = jaccard$\_$similarity(y1,y2)

print (jaccard) \\






$\#$ Cosine similarity\\

def square$\_$rooted(x):

   return round(np.sqrt(sum([a*a for a in x])),3)

  def cosine$\_$similarity(x,y):

 numerator = sum(a*b for a,b in zip(x,y))

 denominator = square$\_$rooted(x)*square$\_$rooted(y)

 return round(numerator/float(denominator),3)  

cosine = cosine$\_$similarity(y1,y2)

print (cosine) \\


$\#$ Pearson correlation\\

my$\_$rho = np.corrcoef(y1,y2)

print (my$\_$rho) \\

$\#$ Levenshtein distance\\

import Levenshtein as lev

ratio=lev.ratio(y1,y2)

jaro = lev.jaro(y1,y2)

jw = lev.jaro$\_$winkler(y1,y2)

ham = lev.hamming(y1,y2)

print (ratio, jaro, jw, ham)

\end{appendix}

\end{document}